\documentclass[prd, onecolumn, nofootinbib, floatfix]{revtex4-1}

\usepackage[mathscr]{euscript}
\usepackage{amsmath}
\usepackage{graphicx}
\usepackage{dcolumn}
\usepackage{bm}
\usepackage{epsfig}
\usepackage{amssymb,latexsym,mathrsfs}
\usepackage{graphicx}
\usepackage{color}
\usepackage{hyperref}
\usepackage{float}
\usepackage{diagbox}
  
\definecolor{vividviolet}{rgb}{0.62, 0.0, 1.0}
\definecolor{amaranth}{rgb}{0.9, 0.17, 0.31}
\definecolor{palatinateblue}{rgb}{0.15, 0.23, 0.89}
\definecolor{brightpink}{rgb}{1.0, 0.0, 0.5}
\definecolor{cornflowerblue}{rgb}{0.39, 0.58, 0.93}
\definecolor{deepcarminepink}{rgb}{0.94, 0.19, 0.22}
\definecolor{radicalred}{rgb}{1.0, 0.21, 0.37}

\hypersetup{ linktoc=all,
    colorlinks, linkcolor={palatinateblue},
    citecolor={brightpink}, urlcolor={amaranth}
}

\newcommand{\be}{\begin{equation}}
\newcommand{\ee}{\end{equation}}
\newcommand{\bs}{\begin{split}} 
\newcommand{\bea}{\begin{eqnarray}}
\newcommand{\eea}{\end{eqnarray}}

\renewcommand{\d}[1]{\ensuremath{\operatorname{d}\!{#1}}}

\begin{document}

\title{Radiation from an inertial mirror horizon}
\author{Michael R.R. Good${}^{1,2}$}
\email{michael.good@nu.edu.kz}
\author{Ernazar Abdikamalov${}^{1,2}$}
\affiliation{${}^1$Physics Department, Nazarbayev University,
\\
Nur-Sultan, Kazakhstan\\
${}^2$Energetic Cosmos Laboratory, Nazarbayev University,\\ Nur-Sultan, Kazakhstan\\
}

\begin{abstract} 
The purpose of this study is to investigate radiation from asymptotic zero acceleration motion where a horizon is formed and subsequently detected by an outside witness. A perfectly reflecting moving mirror is used to model such a system and compute the energy and spectrum.  The trajectory is asymptotically inertial (zero proper acceleration)-ensuring negative energy flux (NEF), yet approaches light-speed with a null ray horizon at a finite advanced time. We compute the spectrum and energy analytically.
\end{abstract} 

\date{\today} 

\maketitle 

Recent studies have utilized the simplicity of the established moving mirror model \cite{moore1970quantum,DEWITT1975295,Davies:1974th,Davies:1976hi,Davies:1977yv,Birrell:1982ix} by applying accelerating boundary correspondences (ABC's) to novel situations including the Schwarzschild \cite{Good:2016oey}, Reissner-Nordstr\"om (RN) \cite{good2020particle}, Kerr \cite{Good:2020fjz}, and de Sitter \cite{Good:2020byh} geometries, whose mirrors have asymptotic infinite accelerations:  
\be \lim_{v\to v_H} \alpha(v) = \infty, \ee
where $v=t+x$ is the advanced time light-cone coordinate, $v_H$ is the horizon and $\alpha$ is the proper acceleration of the moving mirror. These moving mirrors with horizons do not emit negative energy flux (NEF; \cite{Ford:1978qya,Davies:1982cn,Walker:1984ya,Ford:1990id,Ford:1999qv,Ford:2004ba}). ABC's also exist for extremal black holes, including extremal RN \cite{Liberati:2000sq,good2020extreme}, extremal Kerr \cite{Good:2020fjz,Rothman:2000mm} and extremal Kerr-Newman \cite{Foo:2020bmv} geometries, whose mirrors have asymptotic uniform accelerations:
\be \lim_{v\to v_H} \alpha(v) = \textrm{constant}. \ee

These extremal mirrors also have horizons and do not emit NEF. We extend this program by investigating a mirror that has asymptotic zero acceleration:
\be \lim_{v\to v_H} \alpha(v) = 0. \label{inertial}\ee
However, rather than being asymptotically static (zero velocity)  at time-like infinity, $i^+$, \cite{Walker_1982, Good:2019tnf,GoodMPLA,Good:2017kjr,good2013time,Good:2017ddq,Good:2018aer}, which models complete evaporation, or asymptotically drifting (constant velocity) to time-like infinity, $i^+$, \cite{Good:2016atu,Good:2018ell,Good:2018zmx,Myrzakul:2018bhy,Good_2015BirthCry,Good:2016yht}, which model black hole remnants, we seek a globally defined motion that travels to asymptotic {\textit{light-like}} infinity, $\mathscr{I}^+$, forming a finite advanced time horizon, $v=v_H$.  Thus, like the ABC's above with horizons, the mirror we seek needs to travels off to the speed of light, $V\to c$, but in an asymptotically inertial way according to Eq.~(\ref{inertial}).  
Does such an asymptotic inertial horizon-forming mirror exist?  If so, what is the nature of the radiation?
In this note, we answer this question by analytically solving for the quantum stress tensor, beta Bogolubov particle spectrum, and total finite emission of energy for just such a trajectory by the use of an asymptotic inertial {\textit{horizon-forming}} moving mirror.

Our paper is organized as follows: in Sec.\ \ref{sec:motion}, we review the details of the accelerated mirror trajectory, computing only the important but minimal relativistic dynamical properties such as rapidity, speed and acceleration. In Sec.\ \ref{sec:energy}, we derive the energy radiated by analysis of the quantum stress tensor in two different coordinate systems: advanced time light-cone coordinate $v$ and lab Minkowski spatial position coordinate $x$. In Sec.\ \ref{sec:particles}, we derive the particle spectrum, finding a unique Meijer-G form for the radiation and confirm consistency of the results with the stress tensor of Sec \ref{sec:energy}. Throughout we use natural units, $\hbar = c = 1$.
 
\section{Trajectory Motion}\label{sec:motion}
We start with an apriori choice for a (1+1)-dimensional trajectory in space, $x(v)$, as a function of advanced time $v$, that has an obvious singularity at $v=0$:
\be x(v) = -\frac{2 M^3}{v^2},\label{x(v)}\ee
where $v$ is the independent variable, light-cone coordinate, $v= t+x$. Here $M$ is the free parameter scale of the system (not necessarily a mass), which is a positive real constant, $M>0$.
One can immediately see that at the beginning of time, $v\to-\infty$, the position starts off at $x=0$.  As time passes by, $v\to v_H=0$, the mirror heads off to the left (by definition\footnote{We also stress we do not wish to have any run-ins with the mirror; our observer is at $\mathscr{I}^+_R$ and so we require any nascent horizon to be future directed toward $\mathscr{I}^+_L$ avoiding collision and associated divergences in energy flux.}) to $x\to-\infty$.  As is well-known in the literature, e.g. \cite{Ford:1997hb,Hotta:1994ha}, one typically uses light-cone coordinates to express the position as the advanced time position, $p(u)$, with the independent variable, $u$, being retarded time $u=t-x$. Here $p$ as a symbol is used rather than $v$ because $p(u)$ is a function, and $v$ is a coordinate.  

\begin{figure}[H]
\centering
\includegraphics[width=3.5 in]{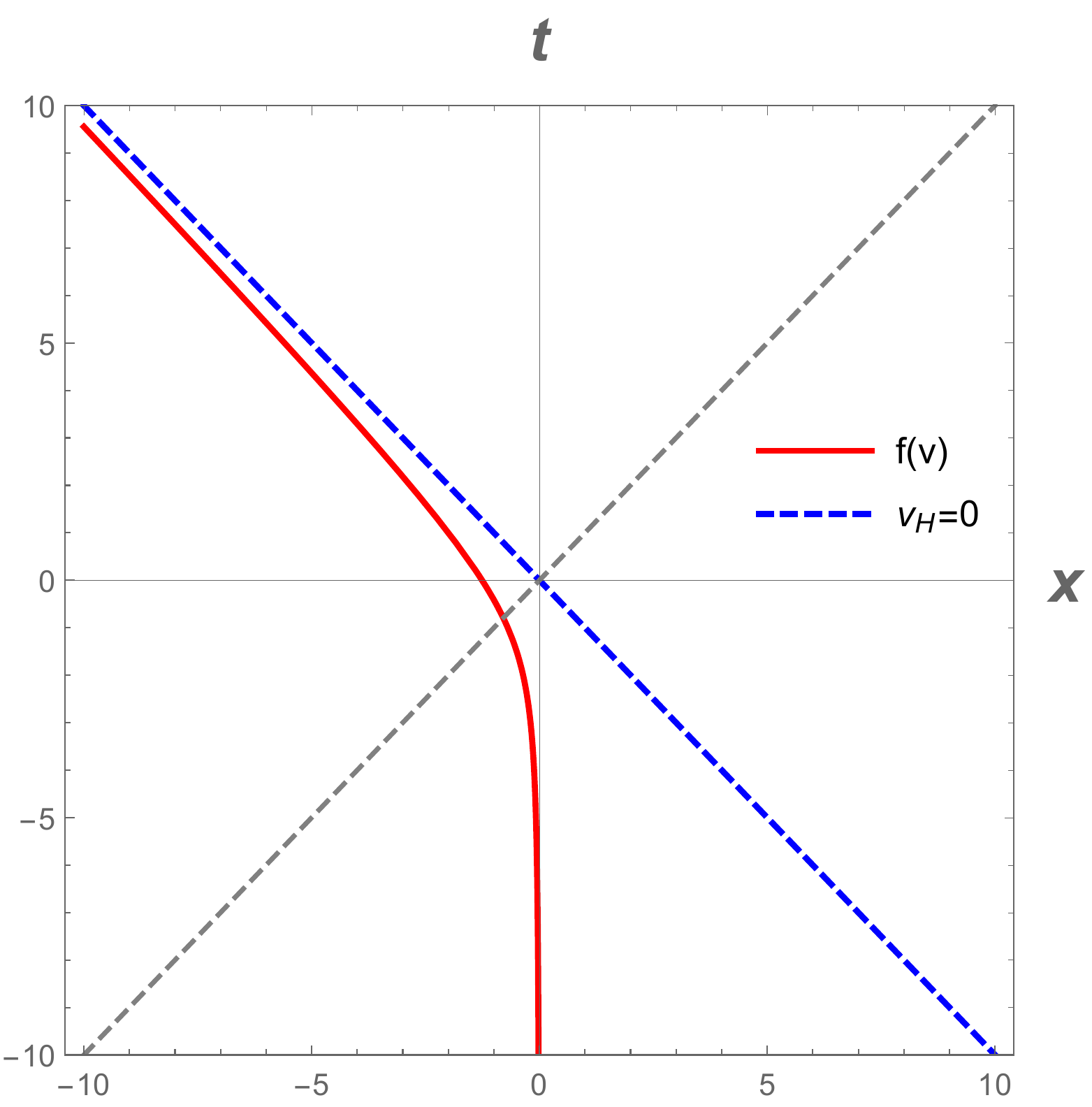}
\caption{A spacetime diagram of the mirror trajectory, Eq.~(\ref{f(v)}). It starts off asymptotically static at $x=0$ with zero acceleration and zero velocity and accelerates to the left eventually reaching the speed of light with a horizon at $t=-x$, or $v_H=0$, evolving to asymptotically inertial motion as $v\to 0^-$. Notice how field modes moving to the left at times greater than $v>0$ will never hit the mirror, demonstrating a bona-fide horizon.  One cannot see any telltale signs from the spacetime graph that $\alpha \to 0$ as $v\to 0^-$.  }\label{Fig1}
\end{figure}

\begin{figure}[H]
\centering
\includegraphics[width=3.5 in]{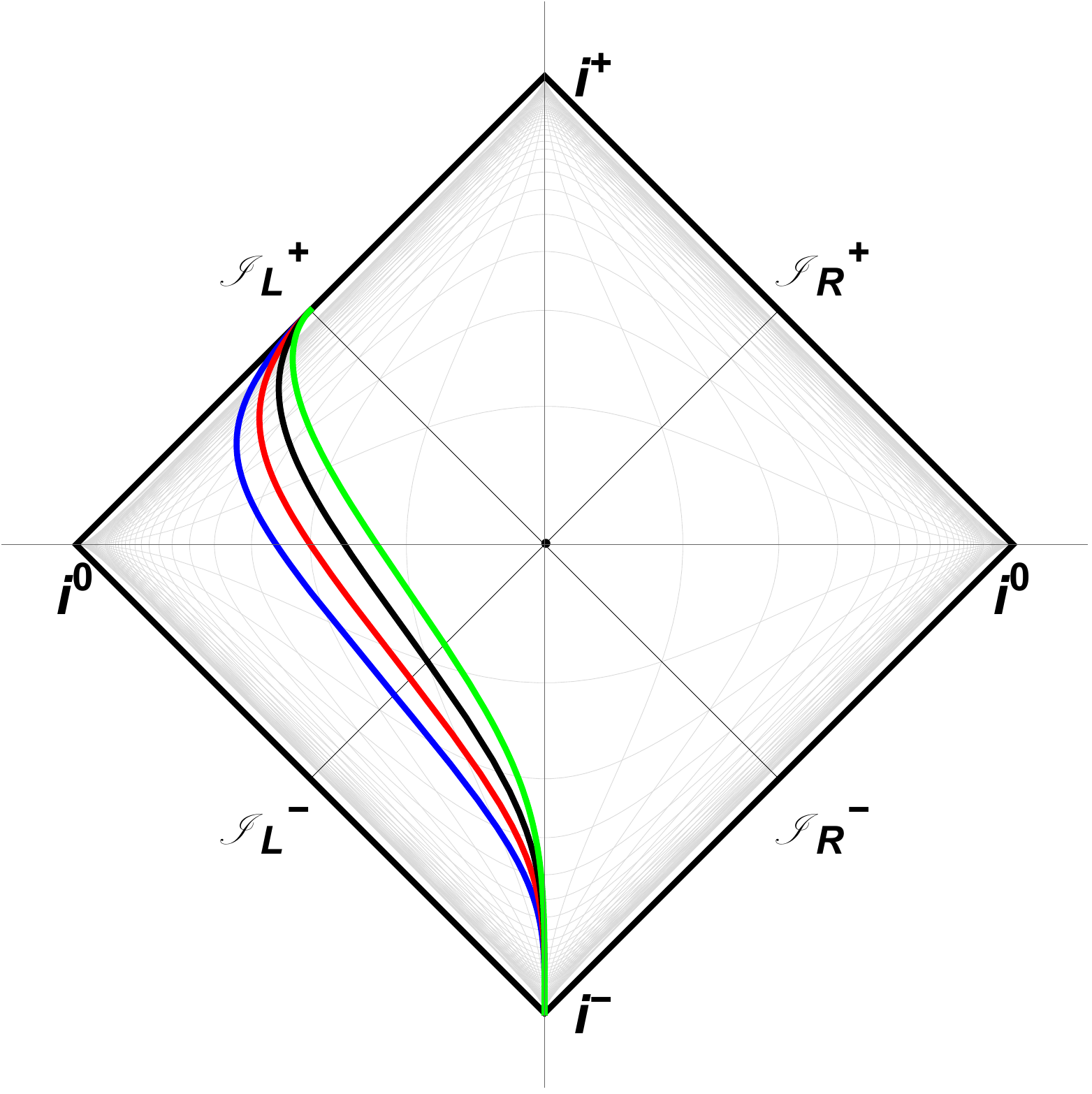}
\caption{A Penrose diagram of the mirror trajectory, Eq.~(\ref{f(v)}). The mirror starts at rest at spatial position $x=0$ when advanced time is $v=-\infty$. It begins to accelerate and as $v\to 0^-$, the velocity approaches the speed of light, and a null horizon forms. The proper acceleration reaches a maximum and then vanishes as $v\to 0^-$.  Since the acceleration is zero as $v\to -\infty$ this this mirror asymptotically inertial.  The various colors correspond to different maximum accelerations.}\label{Fig2}
\end{figure}

For presentation purposes, we would prefer to use $x(t)$ or $p(u)$ but we cannot find a closed-form $x(t)$ or $p(u)$ for our trajectory, Eq.~(\ref{x(v)}), since it is not transcendentally invertible.  Instead, we use light-cone coordinates $(u,v)$, with retarded time $u=t-x$, to express the trajectory Eq.~(\ref{x(v)}) as 
\be f(v) = v + \frac{4 M^3}{v^2},\label{f(v)} \qquad v<0\ee
where $f$ is the retarded time position.  Here $f$ is a function, not a coordinate, so we do not use "$u$", like is the traditional double-duty use of $t$ and $x$ when writing spacetime functions, $[x(t),t(x)]$ and coordinates, $(x,t)$ as independent variables. This has been the common notation since at least the 1970s \cite{Birrell:1982ix}.  For calculations, we will find $f(v)$ to be just about as easy to use as $p(u)$ but a little less intuitive. A spacetime plot with time on the vertical axis is given of the trajectory in Figure \ref{Fig1}. A conformal diagram is plotted in Figure \ref{Fig2}.  Does our trajectory, Eq.~(\ref{f(v)}), incorporate the needed key traits we seek?  
\be
(1)\,\textrm{asymptotic inertial}, \qquad
(2)\,\textrm{asymptotic horizon},
\ee
The horizon (2) can be easily visualized from both Figures \ref{Fig1} and \ref{Fig2}.  However, to confirm that the system is indeed (1) asymptotically inertial, we will need to compute the proper acceleration and relevant limit at $v\to v_H$. 

As a warning, it should be clear that receding at light-speed is not sufficient for a horizon to form.  For example, there are known past moving mirror solutions which are asymptotic time-like and both (1) asymptotic inertial and asymptotic light-speed, but do not form an (2) asymptotic light-like $\mathscr{I^+}$ horizons, \cite{good2013time,Good:2016yht}; i.e., some new proposed solutions may indeed emit NEF but those already known to do so, are not in possession of a horizon.  This highlights the need to proceed with care.
\subsection{Rapidity, Speed, Acceleration}
\begin{figure}[H]
\centering
\includegraphics[width=3.5 in]{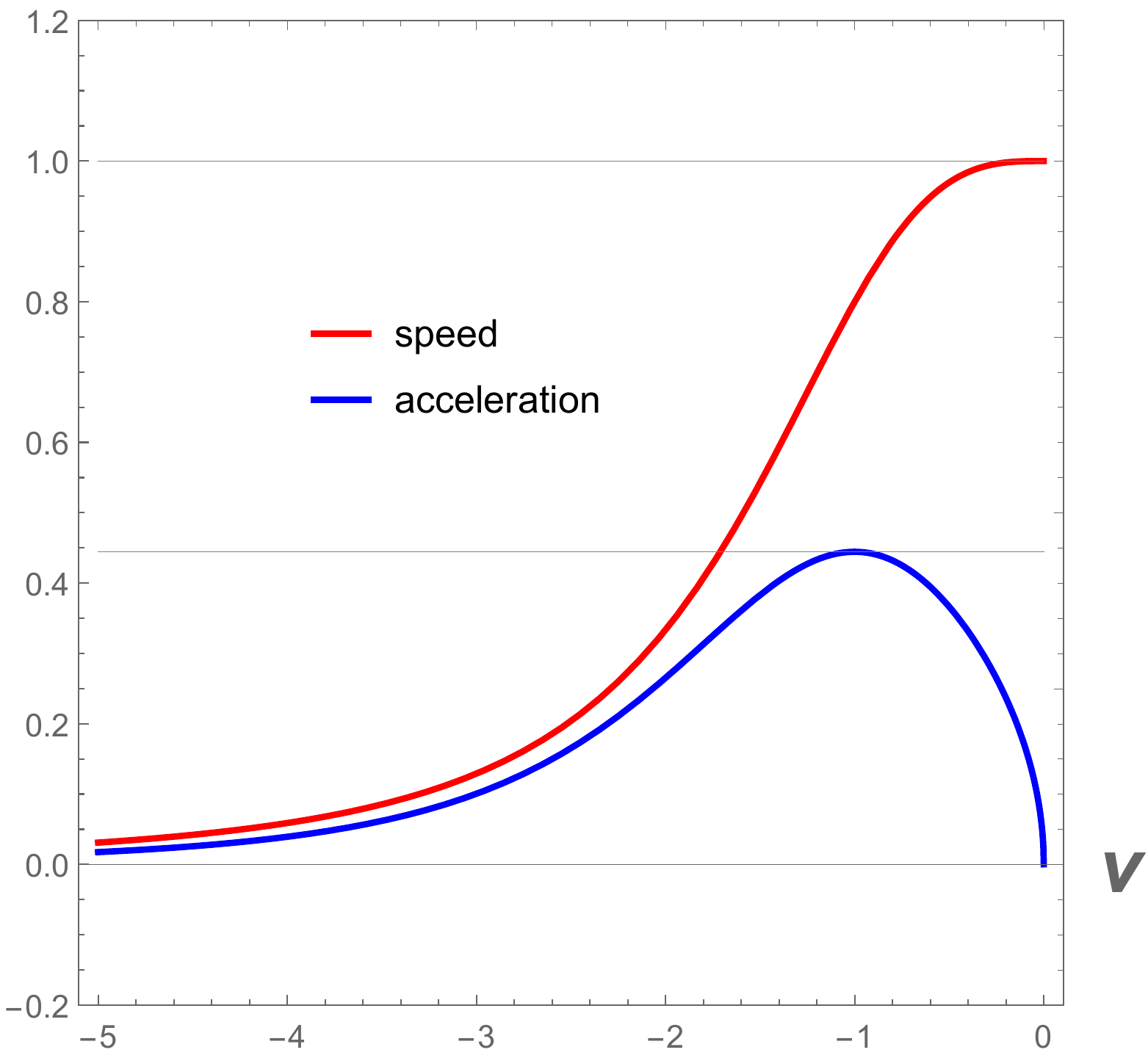}
\caption{Magnitude of the velocity and proper acceleration as a function of light-cone coordinate advanced time $v=t+x$ for the mirror trajectory, Eq.~(\ref{f(v)}).  It is readily seen that at $v=0$, the velocity, $V$, approaches the speed of light, $|V|\to c=1$, and the proper acceleration vanishes, $\alpha \to 0$. The maximum acceleration occurs at advanced time $v=-M$ and is $|\alpha_{\textrm{max}}| = 4/(9M) = 0.444M^{-1}$.  Here $M=1$.}\label{Fig3}
\end{figure}

We compute the rapidity $\eta(v)$, by $2\eta(v) \equiv  -\ln f'(v)$ where the prime is a derivative with respect to the argument, plugging in Eq.~(\ref{f(v)}), 
\be \eta(v) = -\frac{1}{2} \ln \left(1-\frac{8 M^3}{v^3}\right).\label{eta(v)}\ee
With rapidity, we may easily compute the velocity, $V \equiv \tanh \eta$, plugging in Eq.~(\ref{eta(v)}),
\be V(v) = -\tanh \left[\frac{1}{2} \ln \left(1-\frac{8 M^3}{v^3}\right)\right], \label{V(v)}\ee
and the proper acceleration which follows from $\alpha(v)\equiv  e^{\eta(v)} \eta'(v)$, using Eq.~(\ref{eta(v)}) again, so that
\be \alpha(v)= -\frac{12 M^3}{v^4 \left(1-\frac{8 M^3}{v^3}\right)^{3/2}}.\label{alpha(v)}\ee
The magnitude of the velocity, Eq.~(\ref{V(v)}), along with the proper acceleration, Eq.~(\ref{alpha(v)}) are plotted in Figure \ref{Fig3}.  The speed at the start of time, $v\to-\infty$ is immediately seen to be $|V|=0$, while for $v\to 0$, one recedes at the speed of light, $|V|\to 1$, (or $V\to -1$, the negative sign means the mirror is traveling to the left).  The non-monotonic acceleration of Eq.~(\ref{alpha(v)}) is a different story however.  Eq.~(\ref{alpha(v)}) is both asymptotic inertial, with $\alpha = 0$ at the start, $v\to-\infty$ and at the horizon $v\to 0^-$. The maximum acceleration is $|\alpha_{\textrm{max}}| = 4/(9M)$, which occurs at $v=-M$.

\section{Energy Flux and Total Energy}\label{sec:energy}
\subsection{Energy Flux}
\begin{figure}[H]
\centering
\includegraphics[width=3.5 in]{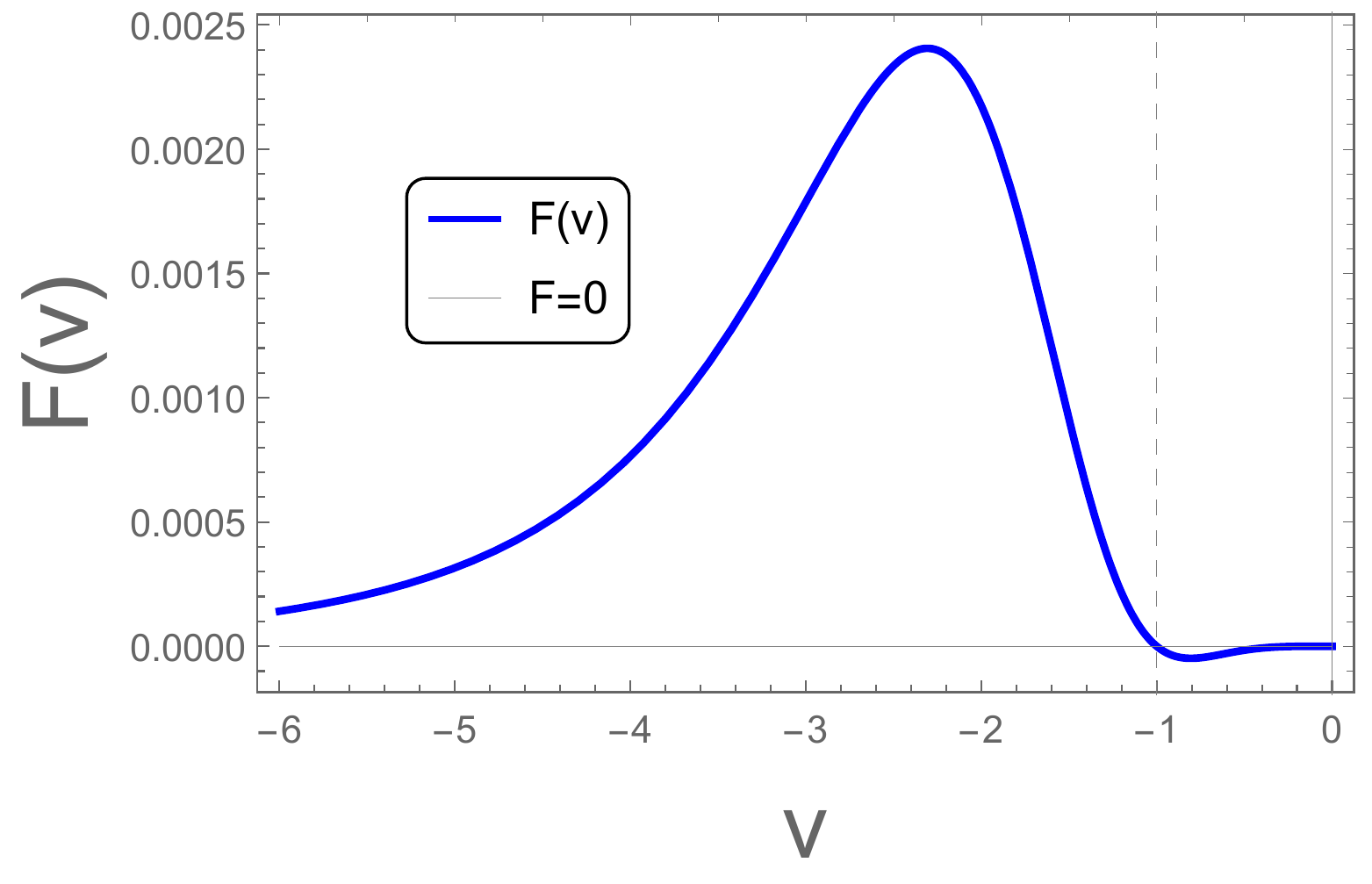}
\caption{The energy flux, Eq.~(\ref{F(v)exact}), is asymptotically zero at $v=-\infty$ and $v=0^-$.  The total energy, as we shall see in Eq.~(\ref{totalE1}), is therefore finite.  Notice the emission of negative energy flux near late advanced times. Here $M=1$.  We have a convenient scale such that at $v=-M$, the energy flux is zero, $F(-1)= 0$. }\label{Fig4}
\end{figure}

\begin{figure}[H]
\centering
\includegraphics[width=3.5 in]{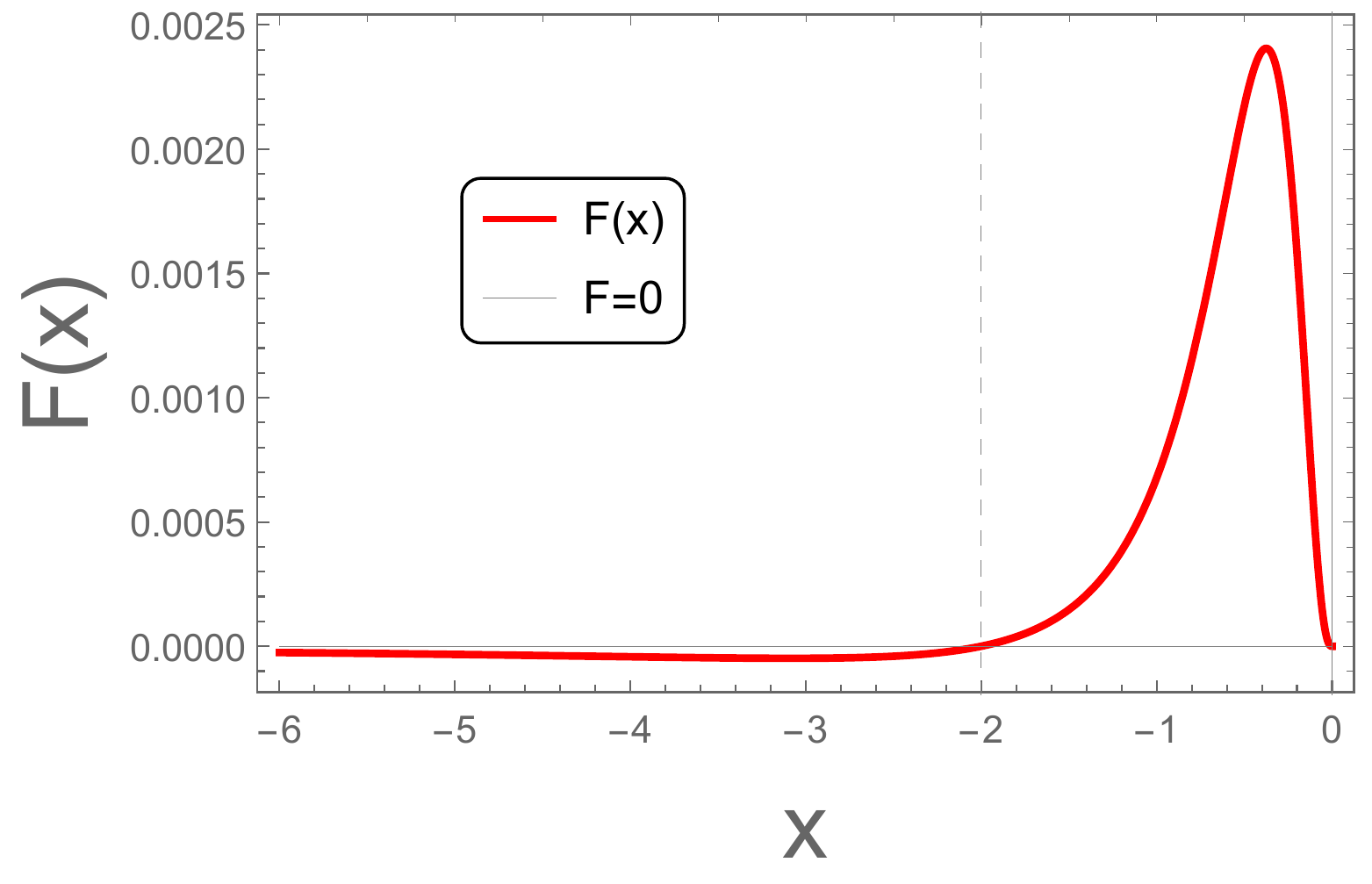}
\caption{The energy flux, Eq.~(\ref{F(x)exact}), is asymptotically zero at $x=-\infty$ and $x=0^-$.  The total energy is finite.  Notice the eternal emission of negative energy flux after $x>-2M$. Here $M=1$. The energy flux is zero, $F(-2)= 0$ and asymptotic approaches zero again as $x\to-\infty$. Note that the mirror starts at $x=0$ and travels to $x=-\infty$, reading the evolution from right to left. }\label{Fig5}
\end{figure}

The quantum stress tensor, $\langle T_{\mu\nu} \rangle$, reveals the energy flux, $F \equiv \langle T_{00} \rangle$, emitted by the moving mirror.  The quantum stress tensor is the vacuum expectation value of the quantized energy-momentum operator in flat space found by point-splitting.  When the mirror has non-uniform acceleration, it is not zero and describes the energy and momentum (not the particles) radiated by the mirror into empty space.  The quantum stress tensor enables one to calculate the rate of energy loss of the mirror. Typically, one sees it expressed as \cite{Davies:1976hi}
\be F(u) = -\frac{1}{24\pi}\{p(u),u\}, \label{F(u)}\ee
where the energy flux, $F(u)$, is a function of light-cone coordinate retarded time, $u = t-x$, \cite{Davies:1977yv, Birrell:1982ix} and the brackets define the Schwarzian derivative. However, since we are using advanced time $v$, we write the radiated energy flux using Eq.~(\ref{f(v)}), \cite{Good:2016atu,Good:2020byh}  
\be F(v)= \frac{1}{24\pi}\{f(v),v\}f'(v)^{-2},\label{F(v)}\ee
where the Schwarzian brackets are defined as usual,
\be \{f(v),v\}\equiv \frac{f'''}{f'} - \frac{3}{2}\left(\frac{f''}{f'}\right)^2\,,\ee 
which gives, using $f(v)= v + 4M^3/v^2$ of Eq.~(\ref{f(v)}), 
\be F(v) = -\frac{4 M^3 v^4 \left(M^3+v^3\right)}{\pi  \left(v^3-8 M^3\right)^4}.\label{F(v)exact}\ee
It is clear that asymptotically $F(v)\to 0$ for both $v\to (-\infty,0^-)$.
A plot of the energy flux, $F(v)$, as a function of advance time $v$ is given in Figure \ref{Fig4}. 
Recall that we used $f(v)$ as the trajectory and $v$ as the independent variable because of analytic transcendental tractability.  That is, we have not found the equivalent trajectory expressed with independent variables proper time $\tau$, retarded time $u$, or coordinate time $t$.  Fortunately, $v$ is sufficient for deriving all the needed and relevant physical observables we are after, i.e. particle production and energy emission. Interestingly, the use of coordinate space $x$ as an independent variable is also readily analytically tractable. It turns out to be illustrative to compute the energy flux as a function of space $x$, in addition to advanced time $v$.  To do this, we express the trajectory Eq.~(\ref{f(v)}) as a timespace function $t(x)$:
\be t(x) = -\sqrt{2} \sqrt{-\frac{M^3}{x}}-x\label{t(x)},\ee
and also express the energy flux as a function of $x$ as well, $F(x)$, with the primes indicative of derivatives with respect to their argument, $x$,
\be F(x) = \frac{1}{12\pi}\left[\frac{t'''(t'^2-1)-3t't''^2}{(t'-1)^4(t'+1)^2}\right].\label{F(x)}\ee
Then, using Eq.~(\ref{t(x)}) and Eq.~(\ref{F(x)}), one finds:
\be F(x)= -\frac{M^6 x^3 \left(2 \sqrt{2} \sqrt{-\frac{M^3}{x}}+x\right)}{\pi  \left(4 x^2 \sqrt{-\frac{M^3}{x}}+\sqrt{2} M^3\right)^4},\label{F(x)exact}\ee
which is plotted in Figure \ref{Fig5}. The benefit of investigating this energy flux in space, is not only an additional mathematical check on the correctness of the results, but an explicit demonstration that negative energy flux is emitted over an infinite distance starting from $x=-2M$ to $x\to -\infty$ in a finite advanced time, $v=-M$ to $v=0$.  

\subsection{Total Energy}
The total energy measured by a far away observer at $\mathscr{I}^+_R$ is \cite{walker1985particle} 
\be E = \int_{-\infty}^{\infty} F(u) \d u, \ee
where integration occurs over retarded time (it takes the energy time to reach $\mathscr{I}^+_R$).  Since we are using advanced time $v$, we write this with $du = \frac{\d f}{\d v} dv$ to get the Jacobian correct,
\be E = \int_{-\infty}^{v_H=0} F(v) f'(v) dv.\label{totalE}\ee
Plugging in Eq.~(\ref{f(v)}) and Eq.~(\ref{F(v)exact}) into Eq.~(\ref{totalE}), a simple analytic form results:
\be E = \frac{1}{72 \sqrt{3} M} = 8.01875 \times 10^{-3} M^{-1},\label{totalE1}\ee
which is finite and positive.  Physically, the finite value tells us the evaporation process stops, similar to the ABC's of extremal black holes (asymptotic uniformly accelerated mirrors \cite{Liberati:2000sq,good2020extreme,Good:2020fjz,Rothman:2000mm,Foo:2020bmv}), black hole remnants (non-horizon asymptotic coasting mirrors \cite{Good:2016atu,Good:2018ell,Good:2018zmx,Myrzakul:2018bhy,Good_2015BirthCry,Good:2016yht}) and complete black hole evaporation (asymptotic static moving mirrors \cite{Walker_1982, Good:2019tnf,GoodMPLA,Good:2017kjr,good2013time,Good:2017ddq,Good:2018aer}). However, it's clear because of the negative energy flux radiated by the horizon at late times, this trajectory models neither an extremal (no NEF), nor a remnant (no horizon), nor complete evaporation (no horizon). The fact that the total energy is positive is consistent with the quantum interest conjecture \cite{Ford:1999qv} as derived from quantum inequalities \cite{Ford:1994bj}.
\subsection{Negative Energy Flux}
Interestingly, as we have seen from Figures \ref{Fig4} and \ref{Fig5}, there is a region of negative energy flux (NEF) at late times or far-left positions.  The NEF begins at $v=-M$ or $x=-2M$. The total amount is
\be E_{NEF} = \int_{v=-M}^{v_H=0} F(v) f'(v) \d v, \qquad \textrm{or}, \qquad E_{NEF} = \int_{x=-2M}^{x=-\infty} F(x) (t'(x)-1) d x,\label{NEFtotal}\ee
which gives an analytic result:
\be E_{NEF} = -\frac{4-\sqrt{3} \pi +\ln 27}{864 \pi  M} = -0.683202\times 10^{-3}M^{-1}.\ee
As a percentage of total energy, Eq.~(\ref{totalE1}), it represents
\be \frac{|E_{NEF}|}{E} \approx 8.52\%,\ee
a fairly substantial amount relative to what it appears to be in Figure \ref{Fig4} with light-cone coordinate advanced time $v$.  Ultimately, this illusionary discrepancy is revealed as such with the inclusion of the Jacobian, $f'(v)$, via integration in Eq.~(\ref{NEFtotal}).
\subsection{Sum Rule for Asymptotic Light-like Horizon}
Let us consider the sum rule of Bianchi-Smerlak \cite{Bianchi:2014qua, Good:2019tnf}, expressed in terms of proper time of the mirror. This sum rule guarantees non-monotonic evaporation via the emission of NEF.  The radiation flux $F(u)$ Schwarzian Eq.~(\ref{F(u)}) is more compactly expressed \cite{Cong:2020nec}:
\be 12\pi F(\tau)= -\eta''(\tau)e^{2\eta(\tau)},\label{F(tau)}\ee
demonstrating that jerking toward an observer at $\mathscr{I}^+_R$, (with positive $+\eta''(\tau)$ jerk), yields negative energy flux.  From Eq.~(\ref{F(tau)}) alone, we know our trajectory Eq.~(\ref{alpha(v)}) will emit NEF.  However, for the sake of argument, let us consider how the sum rule derivation proceeds by integration, which gives   
\be 12\pi\int^\infty_{-\infty} d\tau\,e^{-2\eta(\tau)} F(\tau)=-\left.\eta'\right|^{+\infty}_{-\infty}.\ee
At this point, we can see that any mirror that moves asymptotically with a constant finite rapidity, gives us a zero on the right-hand side because the rapidity $\eta$ as a constant for $\tau \rightarrow \pm \infty$, has derivative zero, and the sum rule is,
\be \int^\infty_{-\infty} d\tau\,e^{-2\eta(\tau)} F(\tau) = 0.\label{sum}\ee
Therefore, on the general principle of a universal asymptotic speed limit for the mirror, that it must always remains time-like (as $\tau\rightarrow \infty$, then $\eta\neq \infty$), asymptotic horizonless mirrors will radiate a negative energy flux.  Through the information-dynamics relationship (e.g. \cite{Chen:2017lum,Bianchi:2014qua}) $\eta = -6S$, the time-like restriction corresponds to a pure state, i.e.  the entanglement entropy, $S$, never diverges and unitary evolution implies negative energy flux; purity $\Rightarrow$ NEF. 

However, clearly the trajectory Eq.~(\ref{f(v)}) is a bit of an outlier (see Table 1), where $\eta \to \infty$ and information loss occurs because $|S|\to\infty$, but NEF is present!  Non-unitary evolution can result in NEF. This is because even though the mirror is not asymptotically time-like, ($\eta$ is never constant, and in fact diverges, $|\eta|\to\infty$ as the mirror goes light-like in finite advanced time), the acceleration, $\alpha = \eta'(\tau)$ still goes to zero at $\tau\to+\infty$. In other words, the asymptotic rapidity, need not asymptote to a finite constant to see NEF radiation. All that is required is that the mirror be asymptotic inertial, which leaves open the strange possibility in the sum rule of Eq.~(\ref{sum}) for a light-like horizon to emit NEF. In this case information loss is coupled with negative energy flux.\footnote{Unitary evolution still implies NEF, regardless.  This conclusion from the sum rule, Eq.~(\ref{sum}), remains unchanged. } 
\begin{table}[H]

\centering
\begin{tabular}{ccccc}
\toprule
\textbf{Mirror Analogs} &
\textbf{Horizon \& Info loss} & \textbf{NEF}	& \textbf{Finite Energy}& \textbf{Finite Particles}\\
Black holes	\cite{Good:2016oey,good2020particle,Good:2020fjz}	        & \checkmark			& X & X & X\\
Extremals \cite{Liberati:2000sq,good2020extreme,Foo:2020bmv,Rothman:2000mm}	            & \checkmark			& X & \checkmark & X\\
Remnants \cite{Good:2016atu,Good:2018ell,Good:2018zmx,Myrzakul:2018bhy,Good_2015BirthCry,Good:2016yht}	            & X			& \checkmark & \checkmark & X\\
Effervescents \cite{Walker_1982, Good:2019tnf,GoodMPLA,Good:2017kjr,good2013time,Good:2017ddq,Good:2018aer}		    & X			& \checkmark & \checkmark & \checkmark\\\\
\textit{Inertial Horizon}	& \checkmark		    & \checkmark & \checkmark & X\\
\toprule
\end{tabular}
\caption{The mirror solution in this paper is fairly unique as measured against previously studied trajectories, incorporating NEF and information loss. In this Table the label `Black holes' refers to Schwarzschild, RN, and Kerr, while the label `Extremals' refer to the extremal RN and extremal Kerr.  The label `Remnants' refer to asymptotically non-zero constant velocity trajectories with a residual field mode Doppler shift, and the label `Effervescents' describe complete evaporation trajectories that are asymptotically inertial with zero-velocity and no residual field mode Doppler shift.}
\end{table}
\section{Particle Spectrum}\label{sec:particles}
\begin{figure}[H]
\centering
\includegraphics[width=3.5 in]{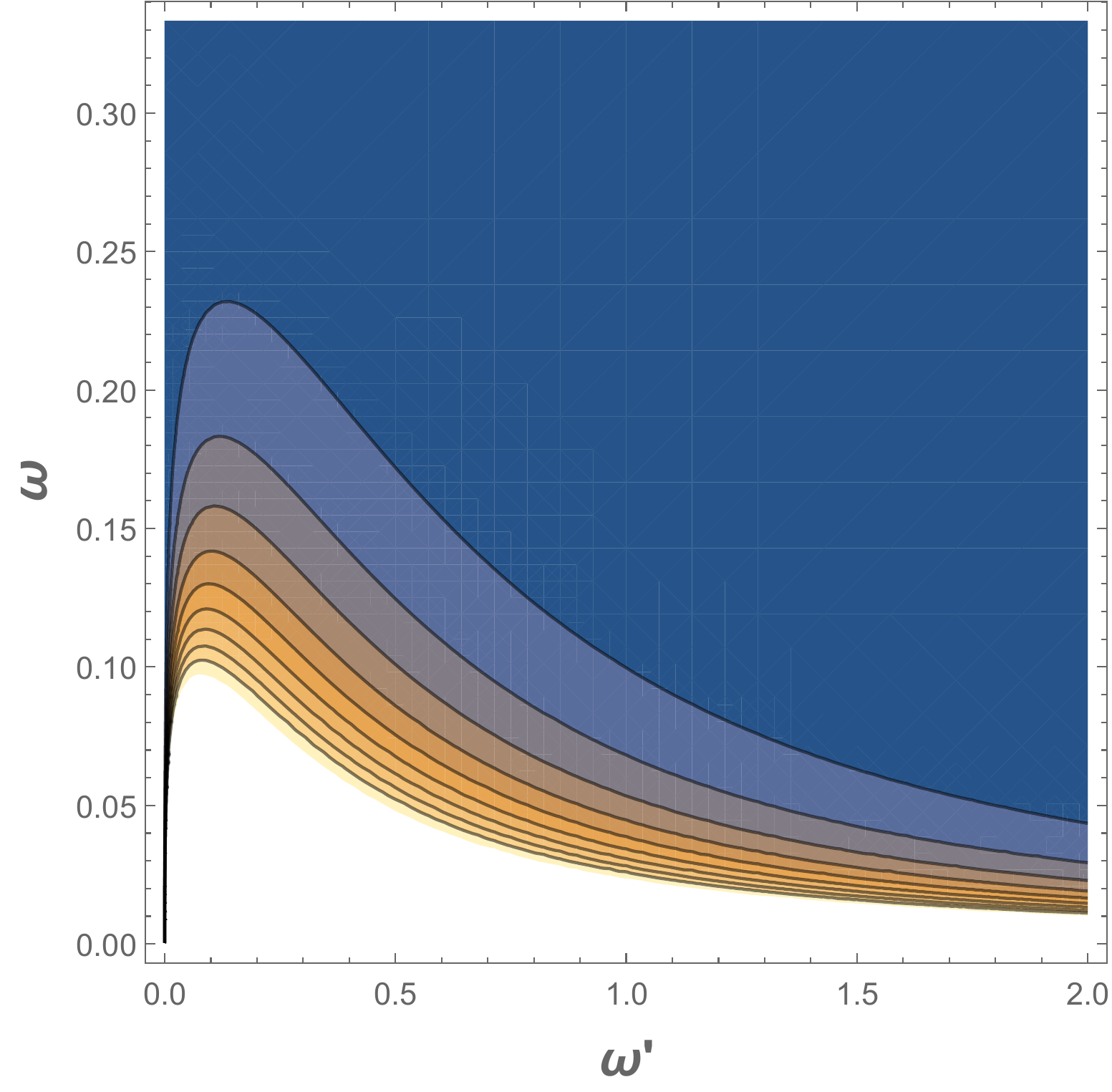}
\caption{The Meijer-G spectrum, Eq.~(\ref{spectrum}), $|\beta_{\omega\omega'}|^2$ as a contour plot, here $M=1$.  Integration over both $\omega$ and $\omega'$ for the integrand $\omega|\beta_{\omega\omega'}|^2$ in Eq.~(\ref{qsum}) gives a finite positive radiated energy consistent with the stress tensor result of Eq.~(\ref{totalE1}). }\label{Fig6}
\end{figure}

The particle spectrum can be obtained from the beta Bogoliubov coefficient, which can be found via \cite{Birrell:1982ix}
\be \beta_{\omega\omega'} = \frac{1}{2\pi}\sqrt{\frac{\omega'}{\omega}}\int_{-\infty}^{v_H} \d v\: e^{-i\omega'v-i\omega f(v)}\,,\label{partsint}\ee
where $\omega$ and $\omega'$ are the frequencies of the outgoing and incoming modes respectively \cite{carlitz1987reflections}. To obtain the particle spectrum, we take the modulus square, $N_{\omega \omega'} \equiv |\beta_{\omega\omega'}|^2$, which gives 
\be  N_{\omega \omega'} = \frac{\omega ' \left|G_{0,3}^{3,0}\left(i M^3 \omega  \left(\omega +\omega '\right)^2|
\begin{array}{c}
 0,\frac{1}{2},1 \\
\end{array}
\right)\right|^2}{4 \pi ^3 \omega  \left(\omega+\omega' \right)^2}.\label{spectrum}\ee
Here $G$ is the Meijer-G function, a general function which reduces to well-known simpler special functions as particular cases.  It's appearance is not unprecedented in studies of the dynamical Casimir effect, e.g. see the effective action of moving mirrors freely-falling onto a black hole \cite{Sorge:2018zfd,Sorge:2019ecb}.
The spectrum Eq.~(\ref{spectrum}), $|\beta_{\omega\omega'}|^2$ is explicitly non-thermal and plotted as a contour plot in Figure \ref{Fig6}. 
\subsection{Limits of the Spectrum}
The leading order term in a series expansion for small $M$ or small $\omega$ is
\be \left|G_{0,3}^{3,0}\left(i M^3 \omega  \left(\omega +\omega '\right)^2|
\begin{array}{c}
 0,\frac{1}{2},1 \\
\end{array}
\right)\right|^2 = \pi.\ee
Therefore the leading order term in small $\omega$ for the spectrum is $N_{\omega\omega'} = (4\pi^2 \omega \omega')^{-1}$, which is identical to the leading order $\omega$ term for the Schwarzschild black hole spectrum at small $\omega$.  However, this similarity between spectrum Eq.~(\ref{spectrum}) and the known spectrum of black holes, Eq.~(\ref{bhspec}), is where the analogy ends.

Looking to late times, which corresponds Hawking's high-frequency approximation $\omega'\gg\omega$ \cite{Hawking:1974sw}, the spectrum is expressed as
\be  N_{\omega \omega'} = \frac{1}{4\pi^3 \omega\omega'} \left|G_{0,3}^{3,0}\left(i M^3 \omega \omega '^2|
\begin{array}{c}
 0,\frac{1}{2},1 \\
\end{array}
\right)\right|^2.\label{spectrumlate}\ee
demonstrating a new form for the late-time spectrum of Hawking radiation emanating from a moving mirror trajectory.  Eq.~(\ref{spectrumlate}) can be compared to the late time spectra of non-extremal and extremal black holes, respectively
\be N_{\omega\omega'} = \frac{1}{2\pi \kappa \omega'}\frac{1}{e^{2\pi \omega/\kappa}-1}, \qquad N_{\omega\omega'} =\frac{1}{{\pi^2\mathcal{A}^2}}\left|K_1\left(\frac{2}{\mathcal{A}}\sqrt{\omega\omega'} \right)\right|^2.\label{bhspec}
\ee
Where $\kappa$ is the surface gravity, i.e. $\kappa = 1/(4M)$ in the case of Schwarzschild, and/or outer horizon surface gravity for the RN and Kerr non-extremal black holes.  Here $\mathcal{A}$ is the extremal parameter, or the asymptotic uniform acceleration \cite{Foo:2020bmv} in the case of the mirror system, where $K_1$ is the modified Bessel function of the second kind with order $n=1$. If $M \gtrapprox \omega$, then a good approximation for the late-time spectrum Eq.~(\ref{spectrumlate}) results:
\be N_{\omega\omega'} = \frac{4 M e^{-3 \sqrt{3} M \sqrt[3]{\omega \omega '^2}}}{3 \pi  \sqrt[3]{\omega ^2 \omega '}}.\ee
The first salient information gleaned from the spectrum is that the inertial horizon is not thermal, i.e. the particles are not in a Planck distribution and there will be no subsequent associated flattening of the energy flux. Second, we can see that the distribution is rather unusual as given by the Meijer-G function, which in this case does not simplify to a more well-known special function. Such a spectrum empirically measured would indicate particle emission from an inertial horizon and a `smoking gun' of sustained negative energy fluxes.

Moreover, the sustained negative energy flux of the inertial horizon trajectory has an interesting relationship to the second law of thermodynamics \cite{Davies:1982cn}.  Since there are no problems with the mirror colliding with the observer, there may be a way to direct an oven into the path of oncoming radiation, cooling its temperature, and lowering its entropy.  Although the beam would create a fluctuation in the total entropy, the second law of thermodynamics being statistical could permit small fluctuations in line with the constraints on NEF \cite{Ford:1990id}. Globally, the strength and duration of the entropy fluctuations would stay safely within permitted statistical boundaries, but locally, a shutter on the oven could be opened at late-times, permitting negative energy flux to gradually accumulate inside. 
\subsection{Does the energy get carried away by the particles?}
The unusual result of the spectrum and rare NEF emission from a horizon emitted to observers at infinity requires checking Eq.~(\ref{spectrum}) for consistency with the total energy, Eq.~(\ref{totalE1}) found from the stress tensor.  This is done by quantum summing,
\be E = \int_{0}^{\infty}\int_{0}^{\infty} \omega N_{\omega\omega'} \d \omega \d \omega',\label{qsum}\ee
that is, associating a quantum of energy $\omega$ with the particle distribution and integrating over all the colors. The result is 
\be E = 8.01783 \times 10^{-3} M^{-1} \approx \frac{1}{72 \sqrt{3} M}.\ee
This result is within $0.01\%$ relative error of Eq.~(\ref{totalE1}), for $M=1$ and the numerical integration results in much better accuracy for larger $M$. It is fine to conclude therefore, that yes, the energy is carried by the particles.  The beta spectrum Eq.~(\ref{spectrum}) is consistent with the quantum stress tensor, Eq.~(\ref{F(v)exact}).  
\section{Conclusions}
In this work, we studied the radiation from a perfect mirror with a particular trajectory coupling information loss and negative energy flux. 
The mirror has a horizon that radiates NEF to an observer at $\mathscr{I}^+_R$.  The motion is ultra-relativistic achieving light speed at a finite advanced time but with asymptotically zero acceleration, leaving the mirror drifting at the speed of light to light-like infinity $\mathscr{I}^+_L$ resulting in the formation of a horizon.  The total energy radiated is positive and the spectrum is that of the Meijer-G function.  Since the energetic radiation is strictly limited by the stress tensor, the evaporation process finishes and the total energy emitted is finite.  Incoming light entering after the horizon has formed is infinitely red-shifted (i.e. never escapes).  This signals the potential existence of a geometry that, like a black hole, has an event horizon that radiates energetic particles to observers at infinity. However at late-times, the horizon of this system is shining with negative energy Hawking radiation.   
\acknowledgments 

MG acknowledges helpful discussions with Eric Linder, Yen Chin Ong, Xiong Chi, Daniele Malafarina, and Joshua Foo.  Funding from state-targeted program ``Center of Excellence for Fundamental and Applied Physics" (BR05236454) by the Ministry of Education and Science of the Republic of Kazakhstan is acknowledged. MG is also funded by the FY2018-SGP-1-STMM Faculty Development Competitive Research Grant No. 090118FD5350 at Nazarbayev University.

\bibliography{main}

\end{document}